\documentclass[fleqn,usenatbib]{mnras}

\usepackage{newtxtext,newtxmath}

\usepackage[T1]{fontenc}
\usepackage{ae,aecompl}

\usepackage{graphicx}	
\usepackage{amsmath}	
\usepackage{amssymb}	

\def\solmass{$M_\odot$}
\def\oiii{[\ion{O}{III}]}

\def\asec{$^{\prime\prime}$}
\def\farcs{\hbox{$.\mkern-4mu^{\prime\prime}$}}
\def\farcm{\hbox{$.\mkern-4mu^{\prime}$}}
\def\kms{km s$^{-1}$}
\def\ha{H$\alpha$}

\title[ULX in NGC 5252]{Stellar Properties of the Host Galaxy of an 
Ultraluminous X-ray Source in NGC 5252}

\author[M. Kim et al.]{
Minjin Kim,$^{1}$\thanks{E-mail: mkim.astro@gmail.com (MK)}
Kristhell M. L\'{o}pez,$^{2,3}$
Peter G. Jonker,$^{2,3}$
Luis C. Ho,$^{4,5}$
and Myungshin Im$^{6}$
\\
$^{1}$Department of Astronomy and Atmospheric Sciences, Kyungpook National 
University, Daegu 41566, Republic of Korea\\
$^{2}$SRON Netherlands Institute for Space Research, Sorbonnelaan 2, NL-3584 CA 
Utrecht, the Netherlands\\
$^{3}$Department of Astrophysics/IMAPP, Radboud University, PO Box 9010, 
NL-6500 GL Nijmegen, the Netherlands\\
$^{4}$Kavli Institute for Astronomy and Astrophysics, Peking University, 
Beijing 100871, China\\
$^{5}$Department of Astronomy, School of Physics, Peking University, Beijing 
100871, China\\
$^{6}$Astronomy Program, FPRD, Department of Physics \& Astronomy, Seoul 
National University, 1 Gwanak-ro, Gwanak-gu, Seoul 08826, Republic of Korea
}

\date{Accepted XXX. Received YYY; in original form ZZZ}

\pubyear{2019}

\begin{document}
\label{firstpage}
\pagerange{\pageref{firstpage}--\pageref{lastpage}}
\maketitle

\begin{abstract}
An ultraluminous X-ray source (ULX) in NGC 5252 has been known as a strong 
candidate for an off-nuclear intermediate-mass black hole. We present 
near-infrared imaging data of the ULX obtained with the William Herschel 
Telescope. Using this data we estimate a stellar mass associated 
with the ULX of $\approx 10^{7.9\pm0.1}$ \solmass, suggesting that it could 
be (the remnant of) a dwarf galaxy that is in the process of merging with 
NGC~5252. Based on a correlation between the mass of the central black hole 
(BH) and host galaxy, the ULX is powered by a $10^5$\solmass\ black hole. 
Alternatively, if the BH mass is $\approx 10^6$ \solmass\ or larger, the host 
galaxy of the ULX must have been heavily stripped during the merger.
The ULX $K_s$-band luminosity is two orders of magnitude smaller than that 
expected from an ordinary active galactic nucleus with the observed \oiii\ 
luminosity, which also suggests the ULX lacks a dusty torus. We discuss how 
these findings provide suggestive evidence that the ULX is hosting an 
intermediate-mass black hole.
\end{abstract}

\begin{keywords}
galaxies: active --- galaxies: individual (NGC 5252) --- galaxies:
Seyfert --- X-rays: galaxies --- black hole physics
\end{keywords}



\section{Introduction}
Supermassive black holes (SMBHs) are nearly always present in the nuclei of 
massive galaxies, and they are thought to play an important role in the 
formation and evolution of the host galaxy. This is for instance inferred from 
the strong correlation between SMBH mass and host galaxy stellar mass 
\citep[e.g. ][]{kormendy_2013}. However, it remains unclear how SMBHs are 
formed and evolve in the early universe. Recent observational studies of high-z 
quasars showed that some SMBHs are already as massive as $10^{8-9}$ \solmass\ 
when the age of the universe was around 0.7-0.8 Gyr 
\citep[e.g.][]{mortlock_2011, banados_2018}. The seed BH mass is a critically 
important parameter which depends on its formation mechanism. It ranges from 
approximately $10^2$ to $10^5$ \solmass\ \citep{volonteri_2010, mezcua_2017}, 
suggesting that intermediate-mass black holes (IMBHs) could be the building 
blocks of SMBHs. However, despite their importance, highly accreting IMBHs are 
rare in the present universe \citep[e.g. ][]{greene_2004}, although IMBHs with 
low accretion rate may be relatively common \citep[e.g. ][]{miller_2015, 
she_2017}. 

\begin{figure*}
	\includegraphics[width=6in]{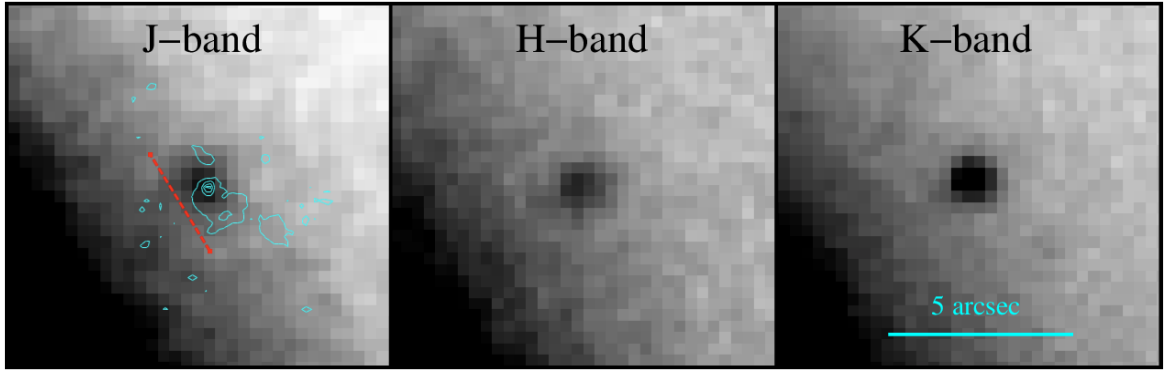}
    \caption{Postage stamp images for CXO~J133815.6$+$043255 in $J$, $H$, and 
$K_s$ from left to right (North is up and East is left). The background 
gradient is due to the stellar light from the host galaxy (NGC~5252). 
Left panel: the overlaid contour represents the \ha\ flux distribution 
which estimated from the {\it HST} image obtained using the 
F673N filter \citep[][]{kim_2015}. The dashed red line denotes the location of
the velocity jump in \ha\ \citep[][]{kim_2017}. The positional 
uncertainties of the WHT and {\it HST} images are approximately 0\farcs15.}
    \label{fig:f1}
\end{figure*}

The possibility to find an IMBH is a strong motivation to study ultraluminous 
X-ray sources (ULX; $L_{\rm X-ray} \geq 10^{39}$ erg s$^{-1}$). 
By definition, the X-ray luminosity of a ULX exceeds the Eddington 
luminosity of a 10 M$_\odot$ stellar-mass black hole, and it does not reside in the nucleus of the apparent host galaxy. 
Observational studies suggest that ULXs are preferentially found in 
low-mass or star-forming galaxies \citep[e.g.][]{swartz_2004, walton_2011}. Such 
high star-formation rate and low metallicity environments are expected 
to form massive stellar BHs efficiently. Therefore, a ULX 
may represent extreme cases of accreting stellar-mass BHs (i.e.~X-ray 
binaries). Theoretical studies also argue that the ULX X-ray luminosity can be 
amplified due to beaming effects \citep{king_2001}, 
further supporting the concept that a stellar-mass BH is the origin of many ULXs.

Furthermore, some ULXs show coherent pulsations indicating it is powered 
by accretion onto a neutron star \citep[e.g.][]{bachetti_2014, 
furst_2016, israel_2017, brightman_2018}.  
Nevertheless, a few ULXs including hyperluminous X-ray sources (HLXs;
$L_{\rm X-ray} \geq 5\times 10^{40}$ erg s$^{-1}$) are 
still regarded as IMBH candidates (e.g. HLX-1; \citealt{farrell_2009}, 
HLX-2; \citealt{jonker_2010, heida_2015}, M82-X1; \citealt{kaaret_2001}, and 
NGC 2276-3c; \citealt{sutton_2012, mezcua_2013}).  

The interesting ULX (CXO~J133815.6$+$043255) was identified in NGC~5252,
a type 2 Seyfert lenticular galaxy. The ULX is 22\asec\ ($\approx10$ kpc) away 
from the centre of NGC~5252 and has an X-ray luminosity $L_{\rm X-ray} \approx 1.2 \times 10^{40}$ erg s$^{-1}$ \citep[][]{kim_2015}. Previous studies using spectroscopic 
data suggest the ULX is kinematically associated with the host with 
velocity offset of $\approx 13$ \kms \citep[][]{kim_2015, kim_2017}. 
Ionized gas appears to be dynamically associated with the ULX, 
again implying that the ULX is not a background AGN \citep[][]{kim_2017}. 
Based on the multiwavelength dataset, the ULX can be explained as an IMBH with a mass of $10^{4-6}$ \solmass\ \citep[][]{kim_2015, mezcua_2018}.

However, the nature of the ULX remains unclear. The host galaxy
does not show morphological signs of an ongoing merger from optical images, although the complex
gas distribution could suggest the host may have undergone a recent minor
merger \citep[][]{prieto_1996}. Therefore, it is natural to suggest that the 
ULX is the remnant core of a dwarf galaxy. The dynamical mass of the ULX counterpart as derived from the 
rotating ionized gas surrounding the ULX, is $\approx 10^{7.5}$ \solmass\, 
supporting such a scenario \citep[][]{kim_2017}. However, we cannot rule out 
that the ULX is a recoiling BH. A key physical parameter to answer this question 
is the stellar mass associated with the ULX. Therefore, in this paper we present near-infrared (NIR) 
photometric data for the ULX CXO~J133815.6$+$043255, and we use it to estimate the 
stellar mass. We then discuss the physical origin of the ULX. 
We assume the luminosity distance to NGC~5252 is 104 Mpc throughout the paper. 

\section{Observations and Data Reduction}
To obtain $JHK_s$ imaging of the ULX we used the Long-slit Intermediate 
Resolution Infrared Spectrograph (LIRIS) mounted on the William Herschel 
Telescope (WHT). LIRIS has a field of view of 4\farcm27 $\times$ 4\farcm27 
and a pixel scale of 0\farcs25/pixel. The $J$-band observations were
obtained using 11 repetitions of an 8-point-dither pattern where 5 images of
20~s exposure each were taken at each point; the $H$-band observations
used 16 repetitions of the same pattern and the $K_s$-band observations
were performed with 22 repetitions of said pattern. The ULX was observed
 in the $H$-band on 2018 March 28, in the $J$-band on 2018 March 29 and 
in the $K_s$-band on 2018 March 30.

Data reduction was performed using the {\scshape theli} pipeline 
\citep{2013ApJS..209...21S}, which produces a master flat in order to 
flat-field correct the data. It also generates a sky background model that is then 
subtracted from each individual image. {\scshape theli} uses {\scshape 
SExtractor} \citet{bertin_1996} to detect sources in each frame and finds an
astrometric solution using {\scshape scamp} \citep{2006ASPC..351..112B}, the 
latter is achieved by matching detected source positions to 
objects from the 2 Micron All Sky Survey (2MASS; \citealt{2006AJ....131.1163S}).
The global astrometric solution is then used for the coaddition of all 
data frames using {\scshape swarp} \citep{2002ASPC..281..228B}. 
The astrometric uncertainty is estimated to be $\sim$0\farcs15, which is 
comparable to that of {\it Hubble Space Telescope} ({\it HST}) images 
\citep{kim_2015}.

The photometric zero-point was defined by comparing the photometric results 
from our imaging data with those from UKIRT Infrared Deep Sky Survey (UKIDSS; 
\citealt{lawrence_2012}) for the same stars. We performed aperture 
photometry for our dataset using the same aperture radius (4\arcsec) as the 
UKIDSS data. Because the UKIDSS data is shallower than ours, we only used 
stars brighter than 18.5 mag for $J$ and $H$ bands and 17.5 mag for the $K_s$-band 
for the comparison. Excluding objects with nearby companions, 10--20 stars 
were finally used. The uncertainties in the zero-point were 0.03, 0.06, and 
0.06 mags in $J$, $H$, and $K_s$, respectively. Throughout this paper, 
all the magnitudes are in the Vega system.

\begin{figure}
	\includegraphics[width=\columnwidth]{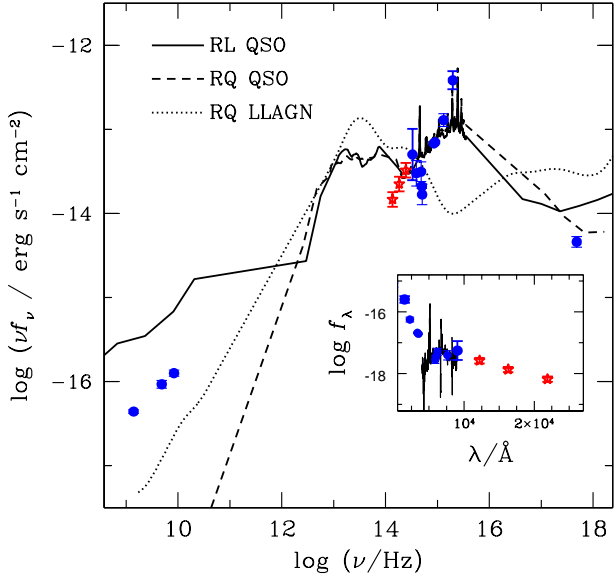}
    \caption{
	The spectral energy distribution for CXO~J133815.6$+$043255. Blue points 
	are from Kim et al.~2015 and references 
	therein; red points are photometric measurements from our NIR images;
	solid, dashed, and dotted lines denote the template SEDs for radio-loud 
	QSOs, radio-quiet QSOs, and radio-quiet low-luminosity AGNs, 
	respectively. Inset: Optical-NIR SED; the solid line represents the optical 
	spectrum of the ULX \citep{kim_2015}; the units on the Y-axis ($f_\lambda$) are
	erg cm$^{-2}$ s$^{-1}$ \AA$^{-1}$.
	}
    \label{fig:f2}
\end{figure}

\section{Photometric Properties}
\subsection{Photometry}
The ULX was clearly detected in all the three bands (Fig.~\ref{fig:f1}). 
Because the ULX resides in the host galaxy outskirt, it is not 
straightforward to measure the ULX brightness. Therefore, we 
performed photometry using three different methods. Aperture photometry with 
an aperture radius of 2 and 4 arcsec provided magnitudes for the NIR 
counterpart of the ULX. These are 
$20.36-20.77$, $19.68-20.01$, and $19.45-19.77$ at $J$, $H$, and $K_s$, 
respectively. Besides, AUTO magnitudes were measured by 
{\scshape SExtractor}, with $J=20.19$, $H=19.81$, and $K=19.44$. Finally, we used 
GALFIT \citep{peng_2002} to properly remove the background gradient caused by 
the host galaxy, modelling the NIR counterpart with a point spread function (PSF). 
We used $\sim$50--100 stars to construct the PSF for each image, yielding 
$J=20.93$, $H=19.97$, and $K=19.52$. Overall, the brightness from the various methods shows a rough agreement within an uncertainty of $0.2-0.3$ mag. These uncertainties were 
considered when we below calculated the error in the stellar mass. We adopt the AUTO 
magnitude in the following discussion.  

We used GALFIT to model the target with an S\'{e}rsic function 
\citep{sersic_1968} to estimate the the NIR counterpart physical size of the ULX. 
However, we found that the 
source was unresolved in all three images. The full-width at half maximum for 
the target ($\sim$ 0\farcs7) was comparable with that for the field stars, 
implying that the target is physically smaller than $\approx$0.3~kpc, which is 
consistent with the finding from the optical counterpart \citep{kim_2015}.

\subsection{The stellar mass and population}
The UV/optical continuum can be dominated by light from accretion onto 
the BH. Thus, the NIR photometry is crucial to estimate the stellar
mass associated with the ULX. We used the $H-K_s$ colour to infer the mass-to-light 
ratio in the $K_s$-band magnitude because those filters are less affected by
emission from the accretion disk. We adopted the conversion factor from 
\citet{bell_2001}, yielding ${\rm log } M_{*,{\rm ULX}} = 7.90\pm0.10$ \solmass. 
The stellar mass derived using $J-K_s$ (${\rm log } M_{*,{\rm ULX}} = 7.25\pm0.10$ 
\solmass) is smaller than that from $H-K_s$. This may indicate that $J$-band is 
somewhat affected by the continuum from the accretion, making the ULX 
bluer (implying a smaller mass-to-light ratio). The IR continuum from a dusty torus 
can partly contribute to the $K_s$-band flux. But we found no clear 
sign of an IR bump from the spectral energy distribution (SED) of the ULX
(Fig.~\ref{fig:f2}), suggesting that the light from 
the torus may be small or even negligible. This is discussed further in \SS{4.2}.
Nevertheless, the stellar mass derived from our NIR photometry can be regarded as an 
upper limit considering a potential contribution from an accretion disk and dusty torus.

Adopting a simple stellar population model from \citet{bruzual_2003},
the observed NIR colour ($H-K_s \sim 0.37$) can be reproduced by old and 
intermediate-age stars ($\sim 10^{8-10}$yr) with supersolar metallicity 
([Fe/H] $\ge 0$). This suggests that the stellar component of the progenitor 
associated with the ULX may arise from a massive system, as inferred from the 
mass-metallicity relation \citep{tremonti_2004}. On the other hand, the $J-H$ 
and $J-K_s$ colours suggest the NIR light could be generated by 
intermediate-age stars ($\sim 10^{8-9}$yr), regardless of their metallicity, 
which would not require the progenitor to be overly massive. It is difficult 
to draw solid conclusions without knowing the exact contribution to the 
brightness from the accretion disk and torus.

\begin{figure}
	\includegraphics[width=\columnwidth]{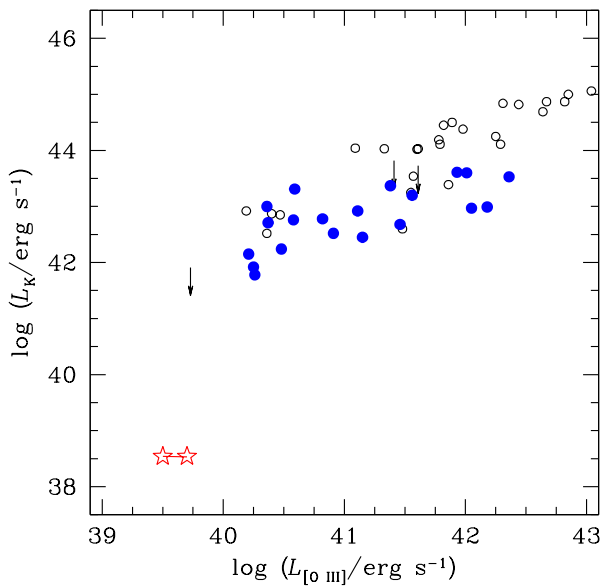}
    \caption{
	Nuclear $K_s$-band luminosity plotted against the \oiii\ luminosity for
	AGNs. Open and filled circles denote type 1 and 2 AGNs, 
	respectively \citep[][]{alonso_1997}. The $K_s$-band luminosity for most 
	targets was measured from the nucleus with $3-6$ arcsec aperture radii. 
	AGNs with upper limit on the $K_s$-band luminosity are shown 
	with arrows. The red stars represent the ULX in NGC~5252. 
	Because the \oiii\ luminosity of the ULX can be overestimated due 
	to the extended features, we set a lower limit on it using aperture 
	photometry with 0\farcs5 radius for IFU data obtained with GMOS 
	\citep{kim_2017}.}  
    \label{fig:f3}
\end{figure}

\section{Off-Nuclear Black Hole Origin}
\subsection{Merging dwarf galaxy?}
Previous studies regarded the ULX in NGC~5252 as an IMBH candidate 
for several reasons. Optical images of NGC~5252 show no evidence for recent 
interaction, suggesting the ULX was less likely to be the nucleus of a massive 
merging galaxy (i.e.~SMBH; \citealt{kim_2015}). The dynamical mass derived 
from gas kinematics in the vicinity of the ULX is $\approx 4 \times 10^7$ 
\solmass\ \citep{kim_2017}. In addition, VLBA observations of the ULX 
showed the BH mass was as large as $10^6$ \solmass \citep{mezcua_2018}. 
For that calculation, the BH mass was inferred from the correlation between 
the X-ray luminosity, BH mass, and radio luminosity (the BH fundamental plane 
of activity, \citealt{merloni_2003}). 
Although the radio luminosity is measured from the radio core, this BH mass 
can be regarded as an upper limit because the contribution from the star 
formation is unknown.

The derived stellar mass (${\rm log } M_* = 7.90$ \solmass) provides 
additional clues for the origin of the ULX. Considering the upper 
limit on the physical size of the stellar component estimated from NIR 
data ($\le 0.3$ kpc), the stellar properties (size and stellar mass) in the 
vicinity of the ULX are comparable to those for ultra-compact dwarfs (UCDs; 
\citealt{fahrion_2019}) rather than conventional bulges.  
Nevertheless, if we assume the ULX is associated with stellar bulges, 
the BH mass is $\approx 10^5$\solmass, using the 
correlation between black hole mass and stellar mass \citep{kormendy_2013}. 
However, there are three caveats 
for this calculation. First, the BH-host mass relation was derived from
the normal galaxies with $M_*>10^9$\solmass, and it is unknown if the 
correlation holds for less massive galaxies \citep[e.g. ][]{davis_2018, 
davis_2019, schutte_2019, woo_2019}. Second, the correlation appears to be tight 
only for elliptical galaxies and classical bulges. For pseudo-bulges, the BH tends to 
be undermassive compared to classical bulges for a 
given bulge mass \citep[e.g. ][]{ho_2014ApJ}. If the ULX was hosted by a pseudo 
bulge, the expected BH mass can be even smaller than $10^5$\solmass.
Finally and most importantly, the stellar component accompanied by the ULX 
could have been tidally stripped by the interaction with NGC~5252, as suggested 
from observational studies of UCDs \citep[e.g. ][]{seth_2014,
ahn_2017, afanasiev_2018}. 
This is also consistent with the apparent supersolar metallicity for 
stars in the vicinity of the ULX estimated from the $H-K_s$ colour. UCDs lies 
above the mass-metallicity relation, suggesting that part of the stellar mass 
of their progenitors was stripped off \citep[e.g. ][]{zhang_2018}. 
The stellar metallicity for massive UCDs (e.g., M60-UCD1 and M59-UCD3) is
comparable to that for the NIR counterpart of the ULX. 

In massive UCDs, the central SMBH mass can be as high as $\sim20\%$ of 
the host stellar mass. Adopting this limit, the BH mass can be up to 
$\approx 10^{6-7}$\solmass, which is in good agreement with the mass derived 
using the BH fundamental plane of activity. If the ULX is the remnant of a 
merging dwarf galaxy, and the BH mass$\approx 10^6$\solmass\ as suggested by 
the BH fundamental plane of activity, the host galaxy progenitor of the ULX is 
expected to be as massive as $\approx 5\times10^8$\solmass. Thus, more than 
30\% of the stellar mass has been stripped off, probably during the merger.     

\citet{kim_2017} concluded that the dynamical mass for this system is $\approx 
4 \times 10^7$ \solmass\ using the ionized gas rotation in the 
vicinity of the ULX. Although this measurement is somewhat uncertain due to the 
unknown inclination angle, it can be regarded as an approximate upper limit 
for the BH mass. Intriguingly, the measured stellar mass is consistent with the dynamical mass. Hence, the BH mass could be significantly 
smaller than the dynamical mass, implying that the ULX can be associated with 
an IMBH. However, the location of the velocity jump is offset from the 
positions of the ULX and NIR counterpart (red dashed line in 
Fig~\ref{fig:f1}). This may indicate that the velocity jump is not due to the 
rotation but instead represents the edge of two ionized gas clouds with 
distinct velocities. Therefore, the dynamical mass calculated assuming the 
sharp transition in the velocity is due to the rotation may not reflect the 
actual mass of the system associated with the ULX.

\subsection{The lack of dusty torus}
The NIR light in AGNs can be emitted not only by stars in the host galaxy, 
but also by a hot dusty torus, and the NIR band contribution from the accretion 
disk may be non-negligible for type 1 AGNs \citep[e.g. ][]{
hernan_2016}. Thus, the $K_s$-band has been widely used to explorer the 
physical properties of torus \citep[e.g. ][]{alonso_1996, koshida_2014}. 
Fig.~\ref{fig:f3} shows a weak correlation between the \oiii\ luminosity and 
the nuclear $K_s$-band luminosity for various AGN types. The nuclear 
$K_s$-band luminosity was measured using aperture photometry, hence the light 
from the stellar components in bulges may be non-negligible. Although the 
scatter is significant, the reprocessed light from the dust is closely related 
to the heating source \citep[i.e. AGN luminosity; ][]{alonso_1997}. The ULX 
lies substantially below the relation of ordinary AGNs\footnote{Considering 
the uncertainty due to the contribution from the extended narrow line region, 
we show the \oiii\ luminosity range in Fig.~\ref{fig:f3}.}. 
This discrepancy is also present in the comparison
of the SED between the ULX and AGNs (Fig.~\ref{fig:f2}), which can be partly 
explained since the contribution from stars around the ULX is 
significantly lower than in the case of Seyferts because the ULX stellar component was heavily 
stripped during the merger. However, this cannot be the whole story as the fraction 
of light from hot dust in the $K_s$-band is known to exceed $20\%$ in type 2 Seyferts. 

The deficit of $K_s$-band flux may indicate that a dusty torus is not present around the ULX. A UV excess is prominent in the SED and no significant absorption was detected in the X-ray spectrum: the ULX is similar in that respect to type 1 AGN. In contrast, broad emission lines are {\it not} detected 
\citep[][]{kim_2015}. The simultaneous absence of evidence for a torus and a broad emission line region 
(BLR) appears to violate the conventional AGN unification model 
\citep[e.g. ][]{antonucci_1993}, suggesting the ULX could be a ``true'' type 2 
AGN, which intrinsically lack a BLR and torus. 
Such type 2 AGNs have either a low Eddington ratio with 
relatively large BH mass \citep[e.g. ][]{tran_2011} or 
high Eddington ratio ($\ge 0.3$) with a low-mass non-stellar BH 
\citep[e.g. ][]{ho_2012, miniutti_2013}.
This can be partly explained by a scenario, where the BLR and torus
were formed from the wind from the accretion disk \citep[e.g. ][]
{elitzur_2009, elitzur_2014}. Based on multiwavelength studies, 
the ULX may host a low mass BH in its centre, hence it is 
likely to have a high Eddington rate (but see \citealt{yang_2017}).  
 

\section*{Acknowledgements}

We are grateful to the referee, Roberto Soria, for very constructive comments.
This research was based on observations made with the William Herschel 
Telescope operated on the island of La Palma by the Isaac Newton Group in the 
Spanish Observatorio del Roque de los Muchachos of the Instituto de 
Astrof\'{i}sica de Canarias. LCH was supported by the National Science 
Foundation of China (11721303, 11991052) and the National Key R\&D Program of 
China (2016YFA0400702). This work was supported by the National Research 
Foundation of Korea (NRF) grant funded by the Korea government (MSIT) 
(No. 2017R1C1B2002879). PGJ and KML acknowledge funding from the European 
Research Council under ERC Consolidator Grant agreement no 647208. KML would 
like to thank Mischa Schirmer for his invaluable help with the data reduction 
pipeline {\scshape theli}. MI acknowledges the support from the NRFK grant, 
No. 2017R1A3A3001362.




\bibliographystyle{mnras}
\bibliography{ulx} 



\bsp	
\label{lastpage}
\end{document}